# Auto-assessment of assessment: A conceptual framework towards fulfilling the policy gaps in academic assessment practices

Wasiq Khan, Luke K. Topham, Peter Atherton, Raghad Al-Shabandar, Hoshang Kolivand, Iftikhar Khan, Abir Hussain

**Abstract** Education is being transformed by rapid advances in Artificial Intelligence (AI), including emerging Generative Artificial Intelligence (GAI). Such technology can significantly support academics and students by automating monotonous tasks and making personalised suggestions. However, despite the potential of the technology, there are significant concerns regarding AI misuse, particularly by students in assessments. There are two schools of thought: one advocates for a complete ban on it, while the other views it as a valuable educational tool, provided it is governed by a robust usage policy. This contradiction clearly indicates a major policy gap in academic practices, and new policies are required to uphold academic standards while enabling staff and students to benefit from technological advancements. We surveyed 117 academics from three countries (UK, UAE, and Iraq), and identified that most academics retain positive opinions regarding AI in education. For example, the majority of experienced academics do not favour complete bans, and they see the potential benefits of AI for students, teaching staff, and academic institutions. Importantly, academics specifically identified the particular benefits of AI for autonomous assessment (71.79% of respondents agreed). Therefore, for the first time, we propose a novel AI framework for autonomously evaluating students' work (e.g., reports, coursework, etc.) and automatically assigning grades based on their knowledge and in-depth understanding of the submitted content. The survey results further highlight a significant lack of awareness of modern AI-based tools (e.g., ChatGPT) among experienced academics, a gap that must be addressed to uphold educational standards.

**Keywords** Education · Assessment · Artificial Intelligence · Machine Learning · LLM · ChatGPT · AI Policy · Academic Policy

## 1 Introduction

AI has tremendously transformed the educational field. Integrating AI technology in the education sector has shaped and fostered the traditional learning method to be more interactive and effective [1]. AI plays a crucial role in supporting higher education by facilitating the learning process for students through providing a personalised learning environment and promoting student engagement [2].

Large Language Models (LLM), such as ChatGPT, have recently demonstrated significant advances in human-like text generation. Their versatility and diverse applications, including question and answering, text classification, text generation, inference, and more [3], have made a significant impact in domains such as public health [4], customer relations [5], finance [6], and education [7]. Despite the promise of the technology, there remains significant controversy regarding its use in academia, with many suggesting it should be banned [8].

Furthermore, the lack of clear policies, guidelines, and frameworks prevents LLMs from being harnessed in academia [9]. Policymakers are struggling to keep up with the rapid development of LLMs and other Artificial Intelligence technologies. This is exacerbated by a lack of understanding of academics' experience and perceptions of LLMs due to insufficient relevant empirical studies [9]. It's crucial that we bridge this gap and ensure that academics' voices are heard in the policy development process to help shape the future of LLM use in academia.

Similarly, most institutions are developing guidelines without collaborating with other institutions. Therefore, there is a lack of consistency in existing policies between institutions [10]. Moreover, a lack of consistency exists between departments, faculties, courses, and even modules within institutions [10]. For example, some institutions encourage module leaders to set the rules and guidelines for LLM use for each assignment [10]. This results in students being allowed to use LLMs in specific assessments, facing restrictions in some assessments and a total ban in others. A lack of consistency results in confusion for academics and students and leads to difficulties in comparing work between modules, courses, and institutions.

Despite LLMs' potential benefits, many academics are concerned with their misuse, such as students submitting generated text as their own during assignments. Such concerns are exacerbated by difficulties differentiating human-written text from LLM-generated text [11]. Such challenges and concerns have led many institutions to ban ChatGPT and other tools [12]. However, such bans are difficult to enforce and result in some students gaining an unfair advantage [12]. Similarly, complete bans prevent students from benefitting from the valuable tools available and pose a barrier to their learning skills, which will benefit them in the workplace [13]. The aforementioned contradictory opinions clearly indicate the major policy gaps in academic institutions that need

3addressing to maintain educational standards across the globe. In this regard, we survey experienced academics to investigate multiple perspectives with the following Research Questions (RQs):

- **RQ1**: Is there a policy gap in students' work assessment within academia in the context of emerging generative AI?
- **RQ2**: Would AI be useful to automatically rank students' work submissions based on their understanding/knowledge of the subject matter rather than solely for detecting ChatGPT (GAI) usage?

The remaining manuscript is organised as follows: the relevant literature is reviewed in Section 2, our methodology, including the data collection method and proposed framework, is described in Section 3, results and discussion and are provided in Section 4, the proposed framework is presented in Section 5, and the conclusion and suggestions for future work are presented in Section 6.

## 2 Related Works

2.1 Proposed Frameworks

Several approaches are suggested to help identify instances where students submit LLM-generated work as their own. For example, [14] suggests:

- Looking for language patterns, irregularities, or grammatical errors, for example, repetitions or low-quality language due to limitations in the LLM.
- Check that references and citations are correct. LLMs are often unable to find genuine references.
- Check for originality. LLMs are incapable of human-level creativity and originality.
- LLMs are known to hallucinate; factual errors may be an indicator.
- Some detection tools are available that automate pattern analysis to identify irregularities.

Although the suggested approaches have shown some success, several limitations affect their practicality. For example, such approaches are time-consuming for already stretched academic staff. Moreover, the suggested approaches are designed to identify LLMs in their current or previous forms. At the current rate of progress, future LLMs will likely render these approaches useless. Even if academics follow such approaches and identify suspected cases of academic misconduct, it is difficult to prove and punish [15].

Alternatively, many institutions suggest altering the style of assessments to make them less susceptible to LLM misuse. For example, [16] suggests interactive assessment activities such as presentations or group discussions to prevent or minimise the potential use of LLMs. Some suggest that such assessment methods may promote independent learning and critical, particularly as there is an opportunity to encourage students to elaborate on or defend specific points [17]. Similarly, "authentic assessments" are promoted, where students are assessed on real-world tasks [17]. Despite the promise of open-ended and authentic assessments, such approaches require significant effort and academic supervision and are not appropriate for all assessments [18].

In summary, most of the proposed frameworks propose methods of reducing the potential for LLM use to minimise the potential for misuse. However, the tools are publicly available and provide many potential benefits, such as improved working efficiency. Therefore, it is logical to assume that workplaces will seek to adopt such tools. Consequently, by preventing students from learning to use such tools effectively, such frameworks may impede students from learning skills that may improve their employability.

### 2.2 Students' Experiences and Perceptions of AI in the Classroom

According to [19], students are generally attracted to AI tools such as ChatGPT and show improved engagement and motivation. However, they suggest that students must be trained to enhance their prompt engineering skills to achieve better results from the tools [19]. Moreover, the need for training is also highlighted [20], who suggested that students are more likely to trust the results provided by AI tools unquestionably. Therefore, they suggest improving students' critical thinking skills and encouraging them to analyse the results' reliability and verify their correctness [20]. Furthermore, in cases where students have been trained to use AI tools appropriately and within the rules of assessment, students have demonstrated a recognition of the value of the ethical use of such tools [21]. However, many students are reluctant to use AI tools as they perceive it as cheating [21].

### 2.3 Academics' Experiences and Perceptions of AI in the Classroom

Teachers also view AI tools as beneficial to their teaching [7], particularly in lesson planning, assessment, and the preparation of learning materials. However, they report concerns regarding the accuracy of information provided, bias, and a lack of human interaction [22]. Similarly, some teachers also highlight the need for students to learn to ethically use AI tools to aid their learning and to prepare them for the future, as they recognise that AI tools are not likely to disappear [23]. Despite many teachers perceiving the ethical use of AI tools as beneficial, many teachers still perceive them as a threat. In addition to the aforementioned academic misconduct concerns, some academics are concerned that using AI tools may decrease students' skills, for example, in analytical writing [24].

### 2.4 Ethics and Authenticity

Ethical frameworks in AIEd tend to be context-dependent and can vary significantly across countries, regions and domains [25]. That said, there are AI policies like UNESCO's, whose global perspective aims to establish adherence to agreed ethical norms [26]. Studies are drawing on growing evidence that LLM can generate high-quality writing for many purposes and in a way that can be passed on to human output. AI writing is hard to detect via anti-plagiarism software and has been shown to be of superior quality to students' own output, including reflective writing [9]. Crompton and Burke's meta-systematic review [27] concluded that higher education uses AI widely for student assessment and evaluation but needs to be managed more ethically, collaboratively and rigorously [27]. Universities are publishing assessment policies in light of GAI, but education is only beginning to allow their policies to be informed by viable conceptual frameworks [28]. While institutions themselves have responsibilities regarding AI and ethics, the prevalence of ChatGPT in students' work also raises questions about students' need to develop a balanced and critical approach to how they use AI [29].

There are ethical concerns in the literature over predictive and GAI. Predictions made by AI can be the product of models that have been trained on biased data [30]. In terms of GAI, while LLM, like ChatGPT, are benefitting students and teachers, there are concerns over bias in both the output of ChatGPT and the broader society. Furthermore, the functionality of some AI has been criticised, for instance, the risk of hallucinations caused by semantic limitations or biased programming [31]. The ethical dimensions of AI Ed require ongoing refinement and augmentation to ensure student and institutional data privacy and minimise bias [32]. Similarly, studies have shown a need for re-skilling educators and addressing the risk of deskilling students [13].

The notion of authenticity is being problematised. Previous studies have perhaps viewed AI as an external assistant to a passive receiver, and this fails to acknowledge the hybrid, collaborative nature of the production of knowledge [28]. AI content, then, has a 'social life' from its conception, production, dissemination, context and multiple uses [33]. Indeed, more recent studies have recommended more human-centred approaches to developing LLMs. Such approaches could acknowledge the synergies between human capabilities and data [34]. In addition to this, a focus on the complementary attributes of humans and AI could be a catalyst for innovation [35]. Moreover, some studies have concluded that AI can make education more human, not less and can also promote well-being if used alongside positive psychology [36]. In counterpoint to this, the literature explores evidence of human intervention having a damaging effect on machine learning [34].

AI policy should focus on ethics, prioritising the development of AI solutions that respect human rights and safety. The development of AI technologies must have a legal and regulatory framework. The aim of this framework is to ensure the transparency and accountability of the AI system [37]. For instance, the governance ethical framework can contain laws and regulations, rules, and ethical committees that focus on ethics, prioritising the development of AI solutions that respect human rights and safety [38].

2.5 Assessment

The assessment of student learning is being augmented by improved AI algorithms, for example, in the domains of predictive learning analytics [7]. AI can aid autonomous learning via students' own questions [39] and assist with answers to open-ended questions [40]. The assessment of students' work can generate real-time data, which AI can evaluate, to help with programme design [41].

Earlier systematic reviews on AI and assessment recognised an absence of discussion of underlying pedagogies that may lead to automated assessment and recommended further research and teacher training [42]. Indeed, ongoing teacher training is recognised as necessary to enable educators to harness new technologies to enhance learning outcomes [43].

Some of the recent literature has examined the architecture of AI Ed on a more granular level, for example, the inaccuracies of static modelling versus the responsiveness and agility of dynamic modelling [7]. Another benefit of dynamic modelling is the accuracy with which systems can alert educators to students who may be at risk of underachieving and, therefore, facilitate timely interventions [7]. More recent literature on LLM builds on studies into predictive modelling techniques from the mid-2010s [44].

This dynamism has informed more recent studies into using LLM to answer open-ended questions, though the range of studies is emerging and the datasets can be sparse [45]. The limited nature of datasets in AI-related studies is a prominent theme in the literature [46]. While some authors may recommend 'massive volumes of standardised datasets' [46], others are more circumspect about the preponderance of big data [43]. Predictive AI can misrepresent students' future behaviours, and the data can be mined as part of a 'digital capitalism', characterised by economic, political, and cultural accumulation and exploitation of data as capital [33].

**3 Methodology**

The methodology proposed in this study comprises two major components:
i) surveying the expert academics and analysing the outcomes to investigate the RQs, and ii) proposing a new framework for the autonomous grading of students' works. For the validation purpose of the proposed conceptual framework, expert opinion is used, drawing on the insights and expertise of individuals who possess in-depth knowledge and experience in the relevant subject area. We surveyed academic experts associated with teaching and learning (T&L) and related administrative roles (e.g., programme leaders, T&L policy managers) to gather information in relation to policy matters and investigate the proposed RQs. Survey questions include awareness, policy on GAI, suggestions, acceptance of proposed AI-based auto-assessment, and other challenges due to the emergence of GAI tools. A detailed list of questions and responses is available in the supplementary material.

Feedback from experts is collected to identify common themes, patterns, and areas of consensus or disagreement, particularly to investigate the proposed RQs. Furthermore, we synthesise experts' opinions to identify key strengths and weaknesses of the proposed conceptual framework in terms of its acceptability and validation for the student assessments.

3.1 Questionnaire

The University Research Ethics Committee (UREC) at Liverpool John Moores University (LJMU) provides ethical approval for the collection and processing of the questionnaire data (Reference: 24/cmp/001). Educational staff, including lecturers, management, and administration staff, were recruited internationally to complete a questionnaire. The questionnaire contained the following questions:

- How familiar are you with ChatGPT, an AI-powered language model used for generating text? (Very familiar / Somewhat familiar / Not familiar)
- Have you encountered instances where students have used ChatGPT or similar AI tools to assist with their coursework or assignments? (Yes / No)
- In your opinion, how has the availability of ChatGPT affected the originality of students' work? (Increased originality / Decreased originality / No noticeable impact)
- How would you rate the overall quality of students' work when ChatGPT is used as a tool? (Improved / Declined / No change)
- How effective is your institution at detecting AI-generated content in students' work? (Very effective / Somewhat effective / Not effective)
- In your opinion, should students be allowed to use AI-generated content in their work/assessments? (Yes / No)
- Does your institution have specific policies or guidelines addressing the use of AI tools like ChatGPT in student work? (Yes / No)
- What strategies, if any, does your institution employ to mitigate the impact of AI tools on assessment integrity? (Check all that apply) (Education on academic integrity / Plagiarism detection software / Manual review of suspicious submissions / Specific guidelines on citation and referencing / Other (please specify))
- To what extent are students aware of the implications of using AI tools like ChatGPT in their coursework? (Highly aware / Somewhat aware / Not aware)
- What challenges, if any, have you encountered in assessing students' work in the presence of AI tools like ChatGPT? (Open text field)
- Do you have any suggestions for improving the assessment process in light of the prevalence of AI tools like ChatGPT? (Open text field)
- Could AI be used to automatically rank students' work submissions based on their understanding/knowledge of the subject matter, rather than solely for detecting ChatGPT usage? (Somewhat aware / Not aware)

- What strategies, if any, does your institution employ to mitigate the impact of AI tools on assessment integrity? (Select all that apply)
- Education on academic integrity
- Plagiarism detection software
- Manual review of suspicious submissions
- Specific guidelines on citation and referencing
- None
- Other (please specify) – Demographic Information: – Role in academia
- Country of work
- Institution type: University, college, school

Table 1: The number and proportions of questionnaire responses for each country.

| Country | Number | Proportion (%) |
|---------|--------|----------------|
| Iraq    | 49     | 43.36          |
| UK      | 40     | 34.19          |
| UAE     | 19     | 16.24          |
| Other   | 5      | 4.42           |

Table 2: The number and proportions of questionnaire responses for each academic job role.

| Role | Number | Proportion |
|------|--------|------------|
| Teaching & Assessment | 105 | 89.74 |
| Management & Policy Matters | 6 | 5.13 |
| Admin | 2 | 1.71 |
| Other | 4 | 3.42 |

3.2 Dataset

A total of 117 responses to the questionnaire were collected. Table 1 displays the number of responses from each country: 43.36% were from Iraq, 34.19% from the UK, 16.24% from the UAE, and the remaining 4.42% were from other countries not listed. Similarly, Table 2 presents an the number and proportions of responses for each academic role. The vast majority are involved in teaching & asessment (89.74%), with the remaining involved in management & policy matters (5.13%), admin (1.71%), or other roles (3.42%).

3.3 Data Processing

The survey outcomes are stored in a secure repository at LJMU. Questions 1 to 9 are multiple choice and, therefore, only accept valid responses via a Microsoft Forms interface, which enforces the relevant rules. Therefore, quality assurance (QA) is automated and preventative. However, the remaining questions contain open fields which require retrospective QA measures. Therefore, open fields were corrected for spelling using spellcheck software. Correcting spelling is necessary for automated analysis, for example, when calculating the number of responses for each country. Furthermore, relevant distribution visualisations were created, such as pie charts to display the numerical proportions of each response to the questions.

## 4 Results and Discussion

4.1 Exploratory Analysis and Visualisation

Table 3 presents the prevalence (i.e., proportion of the respondents as calculated in Equation 1) of each of the possible responses from the MCQs from the survey described in Section 3.1. A total of 117 responses were collected from educational staff in the UK (34.19%), UAE (16.24%), Iraq (43.36%), and others (4.42%). Most responders were primarily involved in teaching and assessment (89.74%), with the remaining responsible for administration (1.71%), management and policy matters (5.13%), and other roles (3.42%).

$$Prevalence = \frac{\text{\# of respondents selecting choice}}{\text{total \# respondents}} \quad (1)$$

Table 3: The prevalence of the multiple choice question responses from the collected survey containing 117 participants.

| Question | Response | Prevalence (%) |
| --- | --- | --- |
| Country of work | Iraq | 43.36 |
| | UAE | 16.24 |
| | UK | 34.19 |
| | Other | 4.42 |
| Role in Academia | Teaching and assessment | 89.74 |
| | Admin | 1.71 |
| | Management and policy matters | 5.13 |
| | Other | 3.42 |
| LLM Familiarity | Very familiar | 43.59 |
| | Somewhat familiar | 46.15 |
| | Not familiar | 10.26 |
| ChatGPT Noticed | Yes | 81.20 |
| | No | 18.80 |
| Originality Affected | Increased | 18.80 |
| | Decreased | 64.96 |
| | No noticeable impact | 16.24 |
| Quality Affected | Improved | 37.61 |
| | Declined | 47.86 |
| | No change | 14.53 |
| Detecting AI-Generated Content | Very effective | 11.97 |
| | Somewhat effective | 42.74 |
| | Not effective | 45.30 |
| Should AI-generated be Allowed | Yes | 54.70 |
| | No | 45.30 |
| Does your institution have policies | Yes | 47.86 |
| | No | 52.14 |
| Student awareness | Highly aware | 22.22 |
| | Somewhat aware | 65.81 |
| | Not aware | 11.97 |
| Institutional strategies | Education on academic integrity | 53.85 |
| | Plagiarism detection software | 60.68 |
| | Manual review of suspicious submissions | 47.86 |
| | Specific guidelines on citation and referencing | 39.32 |
| | None | 8.55 |
| | Other | 1.71 |
| Use of AI for auto-assessment | Yes | 71.79 |
| | No | 28.21 |

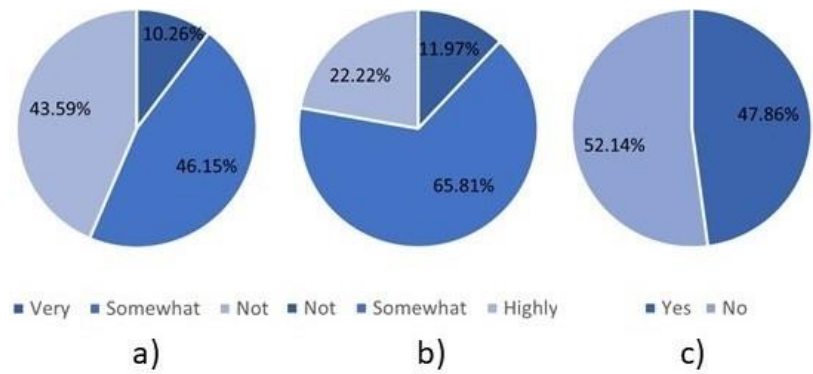

Fig. 1: Pie charts demonstrating the limitations in terms of the awareness of AI in education. a) LLM familiarity b) student awareness c) Institutional policy

Also shown in Table 3 and Figure 1, there are important limitations regarding the awareness of AI and relevant educational policies. For example, only 43.59% of responders stated that they were "very familiar" with LLMs, with the remaining "somewhat familiar" (46.15%) and "not familiar" (10.26%). Similarly, only 22.22% of responders stated that their students were "highly aware" of the implications of using AI tools in their coursework, with the remaining "somewhat aware" (65.81%) and "not aware" (11.97%). Moreover, it is likely that students' lack of awareness may be related to the fact that only

47.86% of responders reported that their institutions had policies regarding the use of AI in assessments. Such results suggest a need for improved awareness for both staff and students in addition to formal policies to ensure that staff and students are aware of the rules and regulations surrounding the use of AI in education.

Moreover, Table 3 and Figure 2 further highlight the importance of AI awareness and policies. For example, 81.20% of responders reported that they had identified evidence of AI usage in student assessments. Moreover, 64.96% of responders reported that AI usage reduced assessment originality, and 47.86% also stated that it negatively affected the quality of student work. Furthermore, as shown in Figure 3 c), only 11.97% of responders reported that their institution was effective at identifying AI-generated content in student work, with the remainder stating that their institutions are somewhat effective (42.74%) or not effective (45.30%). These are concerning statistics as they suggest that academic standards may be threatened and institutions may not be well-equipped to handle the current situation.

Furthermore, Table 3 and Figure 3 describe academics' views regarding the use of AI in education. Despite the aforementioned issues, academics tend to have positive opinions of AI in education. For example, 54,70% of responders stated that they believed that AI should be allowed in assessments. Similarly, 71.79% of responders indicated that they believed that AI-based autonomous assessment would benefit them.

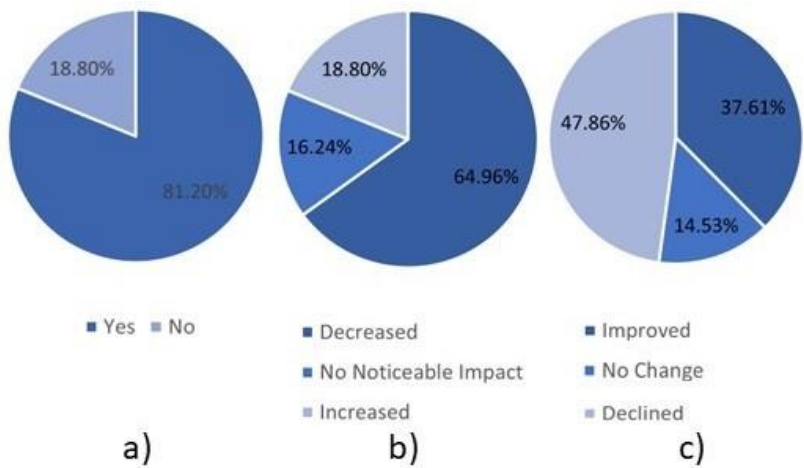

Fig. 2: Pie charts demonstrating the impacts of AI in education. a) AI identified b) effect on originality c) effect on quality

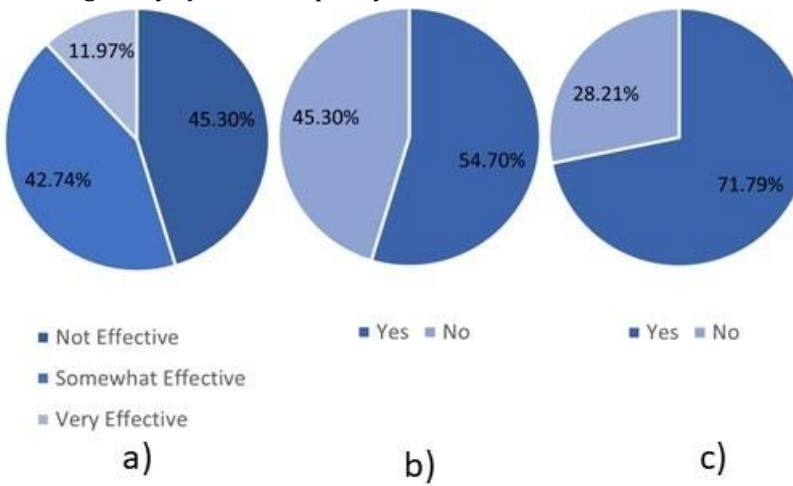

Fig. 3: Pie charts describing academics' opinions regarding AI in education.
a) identification effectiveness b) should AI be allowed c) would autonomous assessment be beneficial

4.2 Statistical Analysis

Chi-square Test: The Chi-square test is an established statistical method to determine the significance of dependence between two categorical variables. In this study, we utilise the Chi-square test to statistically determine whether the acceptability of AI-based tools for the auto-assessment, as well as the use of AI-based tools by the students, are associated with other factors (e.g., familiarity with AI, institutional policy, etc.). The outcomes from a Chi-square test primarily include:

- **Chi-square statistic ($\chi^2$):** This test statistic is calculated based on the differences between observed and expected frequencies. A higher value indicates a greater divergence between the observed and expected data.
- **p-value:** The probability that the observed differences occurred by chance. If the p-value is less than a chosen significance level (we used the standard one: 0.05), it suggests that the difference between observed and expected values is statistically significant.
- **Degrees of Freedom (df):** This is based on the number of categories in the variables

Table 4: Interdependence between the recommendation of AI-based auto assessment tools, acceptance of the usage of AI-based tools by students, and other related factors.

|  | GAI for auto-assessment | | | Allow GAI | | |
| --- | --- | --- | --- | --- | --- | --- |
|  | $\chi^2$ | df | *p*-value | $\chi^2$ | df | *p*-value |
| LLM-Familiarity | 0.39 | 2 | 0.82 | 7.7 | 2 | 0.02 |
| ChatGPT noticed | 0.46 | 1 | 0.49 | 6.91 | 1 | 0.008 |
| Originality affected | 5.35 | 2 | 0.06 | 12.02 | 2 | 0.002 |
| Quality affected | 2.14 | 2 | 0.34 | 15.68 | 2 | 0.0003 |
| Detecting AI-generated content | 0.39 | 2 | 0.81 | 0.60 | 2 | 0.73 |
| Allow-GAI | 7.31 | 1 | 0.006 | 7.31 | 1 | 0.006 |
| Policies | 0 | 1 | 1 | 0.48 | 1 | 0.48 |
| Awareness | 1.52 | 2 | 0.46 | 5.09 | 2 | 0.07 |

Table 4 shows the statistical significance of dependence between various factors (i.e., questions asked) and experts' opinions on the acceptability of GAI for students' work submission and the use of GAI for the automated assessment (i.e., proposed framework). It can be noticed that the acceptability of GAI is significantly dependent on most of the factors (i.e., p¡0.05) except detecting AI-generated content (p=0.07), policies (p=0.48), and awareness (p=0.07). It can be noticed that the acceptability of GAI for students' work submissions is significantly dependent upon familiarity with AI-based tools (p¡0.02). The survey outcomes show that 67% of the responses with high familiarity with LLM showed acceptance of the use of LLM by the students. On the other side, it decreased to 25% only in the case of participants who are unfamiliar with AI-based tools.

In regard to the use of GAI for automated grading and assessment, no significant relationship is noticed mostly (p>0.05) except 'allow-GAI', which indicates strong interdependence (p=0.006). In other words, regardless of familiarity, awareness, and other factors (Table 4), the use of AI-based autonomous assessment (i.e., proposed framework) is recommended by the participants. Furthermore, the distribution of responses indicates that a high proportion (83%) of the staff who accept the use of AI in students' works also recommend using AI-based tools for automated assessment (i.e., proposed idea). However, this opinion decreases to 58% for participants who believe students should not use GAI in their submission.

These statistics clearly support the argument set in RQ2 that AI-based tools would be useful for the automated grading of students' work submissions based on their understanding and knowledge of the subject matter instead of detecting ChatGPT (GAI) usage (which is common practice everywhere due to policy gaps). The statistics also support the concept of allowing students to utilise GAI for the assessments (e.g., report writing, coursework, etc.) while new policy standards would be required to do so.

As shown in Table 3, 81.20% of the academic staff surveyed reported noticing GAI in students' assessments. However, only 10.26% of those surveyed reported that they were very familiar with LLMs. Therefore, it is clear that training is required in established detection techniques, such as those described in [47].

Similarly, only 11.97% stated that students were highly aware of the implications of GAI in assessments. Recent works such as [48] have investigated students' perceptions of AI usage and have found that they are popular. However, they have not investigated the awareness of potential implications such as academic misconduct. Therefore, there is a clear need to ensure that all students are familiar with current policies and the implications of unethical GAI usage.

Despite the prevalence of GAI, many academics consider a complete ban of GAI necessary to maintain academic standards [49]. However, the questionnaire results suggest that both academics and students find benefits from such technology, benefits that would be lost under a total ban. Moreover, existing works such as [48] suggest that students can use tools such as ChatGPT ethically to support their studies.

However, it is clear that there is a lack of consistent policy regarding AI usage in academia, particularly surrounding assessment. As shown in Figure 4 there is no clear single policy. A broad mixture of policies are reported, and 10% of respondents report that no policy is employed in their institution. Similarly, as reported in Table 3, only 47.86% of respondents stated that their institution had specific policies related to AI usage. This is supported by works such as [50], which highlight a lack of clear and consistent policies. Therefore, there is a clear need for consistent policies in academia. However, such policies must be robust and promote the ethical use of AI. As shown in Table 3, 54.70% of academic respondents favour allowing students to use AI. Therefore, it is unlikely that academics would desire a complete ban.

Moreover, academics also see the potential benefits of AI for themselves. For example, Table 3 shows that 71.79% of respondents would benefit from assessing students and providing feedback. Therefore, there is evident demand for an automated assessment framework like that we propose in Section 5. Hence, future work will aim to implement this.

4.3 Open-Ended Question Analysis

The open-ended questions described in Section 3 were analysed using Latent Dirichlet Allocation (LDA) [51] to identify common topics and keywords. Two open-ended questions are provided in Section 3. The first is related to challenges that academics have faced, and the second asks academics to provide suggestions to improve assessments in the presence of AI tools.

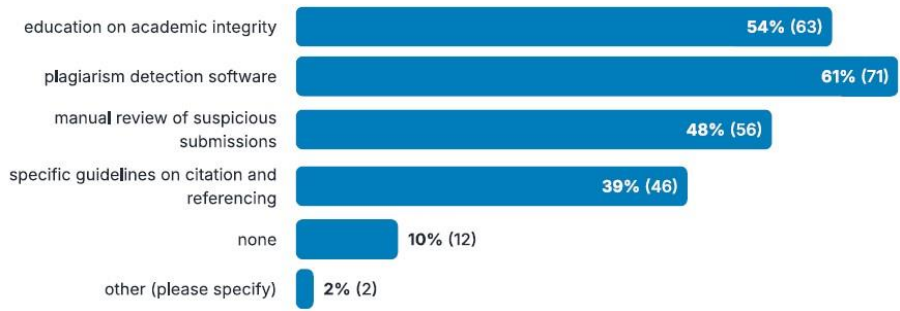

Fig. 4: Responses to the question "What strategies, if any, does your institution employ to mitigate the impact of AI tools on assessment integrity? (Check all that apply)"

*What challenges, if any, have you encountered in assessing students' work in the presence of AI tools like ChatGPT?*

In response to the aforementioned question, several keywords suggest common challenges. For example, "detection", "tool", "understanding", and "originality" are amongst the most common keywords. Identifying the common keywords in the original responses highlights common challenges, such as a lack of originality, as already highlighted in Table 3. Additionally, it highlights additional concerns related to assessing students' understanding in the presence of tools such as ChatGPT. Similarly, many academics wish for better tools to detect GAI usage.

Do you have any suggestions for improving the assessment process in light of the prevalence of AI tools like ChatGPT? A plethora of keywords are presented in response to the request for suggestions. For example, common keywords include "guideline", "policy", "viva", "presentation", "face [to face]", "written", "practical", "diversify", and "exam". Moreover, further investigation of the suggestion shows a common theme of alternative assessment. For example, many responses suggest alternative and diverse assessments to prevent the use of AI in assessments, such as presentations, oral responses, exams, and practicals. Another common suggestion is to create and implement clear AI guidelines and policies.

## 5 Proposed Framework

As described previously, 71,79% of respondents reported that they would approve of the use of AI for auto-assessment. Therefore, in Figure 5 we present a proposed framework for automatically assessing students' work. Students' original work submissions (e.g., code, reports, etc.) will be forwarded to the GAI model (e.g., ChatGPT). The model will randomly generate N number of Multiple Choice Questions (MCQs) from the submitted work along with the associated correct answers. The auto-generated MCQs are then forwarded to institutional repositories or Virtual Learning Environments (e.g., Canvas, Blackboard, etc.), where students will be given a preset timeline to respond to the MCQs. These responses will then be automatically compared with the auto-generated correct answers to produce a grade for the corresponding work. Descriptive feedback will then be generated based on the student's responses and the scores they achieved. Then, the usual administrative and quality control processes are followed, such as assessment moderation and the delivery of grades and feedback to students.

---

**Algorithm 1** Proposed automated assessment of assessments.

---

*Receive Student Submission*
**Input**: Original student work (code, report, etc.)
**Action**: Student submits work through a designated portal or VLE. **Output**: Submission is forwarded to the GAI model.

*Generate MCQs from Student Submission* **Input**: Student's submitted work.
**Action**: The GAI model (e.g., ChatGPT) analyses the submission. **Process**:

- Extract key concepts from the submission.
- Randomly generate N number of Multiple Choice Questions (MCQs) based on these concepts.
- Generate correct answers for the MCQs.

**Output**: A set of N MCQs and their correct answers.

*Forward MCQs to Institutional Repository/VLE* **Input**: Generated MCQs and answers.
**Action**: The MCQs are uploaded to the VLE (Canvas, Blackboard, etc.). **Output**: Students access the MCQs within a preset timeline.

*Collect Student Responses*
**Input**: Student responses to the MCQs within the VLE.
**Action**: Students submit their answers.
**Output**: Responses are stored for comparison.

*Compare Responses with Correct Answers*
**Input**: Student responses and auto-generated correct answers.
**Action**: Responses are automatically compared with the correct answers using a predefined grading algorithm.
**Output**: Raw grade based on correct/incorrect answers.

*Generate Descriptive Feedback*
**Input**: Student responses and achieved scores.

**Action**: The AI model generates feedback based on the following: Correctness of answers.

Similarly, Algorithm 1 describes the logic relating to the proposed automated assessment of assessments. The process begins when students submit work via existing methods, such as a VLE. The submission is then forwarded to a GAI model. The GAI model then generates MCQs from the submitted work by extracting key concepts from the submission. Likewise, correct answers are generated for the MCQs. The MCQs can then be sent to the students via an institutional repository or VLE. Students may then submit their answers, which are compared to the correct answers for grading. Lastly, the GAI model generates feedback based on the returned answers and the correctness of the answers.

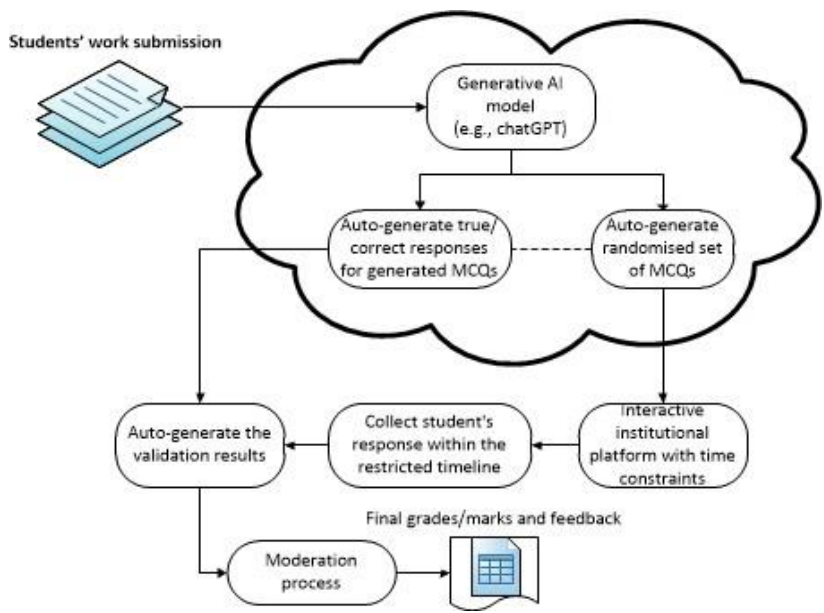

Fig. 5: Proposed conceptual framework for the auto-assessment of students' work submission comprising GAI, rule-based algorithm, and human intervention.

## 6 Conclusion and Future Work

The results described in Section 4 suggest that staff and students lack sufficient awareness of AI in education. This is exacerbated by the fact that more than half of institutions do not provide relevant policies related to AI used in assessments. Moreover, it is apparent that many institutions cannot identify instances where students submit AI-generated content in their assessments.

Therefore, there is a clear need for training, identification tools, and the development and implementation of relevant policies.

However, despite the aforementioned issues surrounding AI-generated content in assessment, as described in Section 4, academics tend to remain optimistic regarding the use of AI in education. In particular, the vast majority of responders (71.79% prevalence) stated that they would benefit from AI tools such as autonomous assessment (RQ2). Therefore, we propose the conceptual framework provided in Figure 5 to meet these needs.

Furthermore, as described previously, the literature [8] reports many instances of AI tools being banned or restricted in academic policies. However, the positive opinions reported in Section 4 suggest that such policies may be unpopular with academics. Similarly, it indicates a policy gap concerning students' assessment in the context of emerging GAI (RQ1). Similarly, responses to the open-ended responses reported in Section 3 suggest that academics strongly desire clear and consistent policies and guidelines. Therefore, further consultation with academics is suggested to develop policies that balance upholding academic standards and encouraging students and academics to learn the current technology and reap the benefits [49].

Similarly, examining the open-ended responses reported in Section 3 suggests that a ban may not be necessary. For example, many responses suggested that students could be prevented from using AI tools in assessments by diversifying in terms of the types of assessments requested of students. Alternative assessment methods such as presentations, oral presentations, practicals, and exams are unlikely to be aided by the misuse of AI.

Future work should be performed in collaboration with academics, policymakers, and students to understand better what the stakeholders want and need in updated academic policies. Such policies should adapt to the everevolving technology landscape and uphold academic standards without rejecting the potential benefits. Furthermore, future work is required to implement the automated assessment tool described in 5. As evidenced in 4, such a framework is desired by academics.